\documentclass[journal=jacsat,manuscript=article]{achemso}

\usepackage{chemformula} 
\usepackage[T1]{fontenc} 

\author{Ygor M. Jaques}
\affiliation{Group of Organic Solids and New Materials (GSONM), Gleb Wataghin Institute of Physics, University of Campinas (UNICAMP), Campinas, SP, Brazil.}
\alsoaffiliation{Center for Computational Engineering \& Sciences (CCES), University of Campinas - UNICAMP, Campinas, SP, Brazil.}
\author{Cristiano F. Woellner}
\affiliation{Department of Physics, Federal University of Parana UFPR, Curitiba, Paraná 81531-980, Brazil}
\author{Lucas M. Sassi}
\affiliation{Materials Science and Nanoengineering, Rice University, Houston, Texas 77005, USA.}
\author{Marcelo L. Pereira Jr}
\affiliation{University of Bras\'{i}lia, College of Technology, Department of Electrical Engineering, Bras\'{i}lia, Federal District, Brazil.}
\alsoaffiliation{Materials Science and Nanoengineering, Rice University, Houston, Texas 77005, USA.}
\author{Luiz A. Ribeiro Jr}
\affiliation{Computational Materials Laboratory, LCCMat, Institute of Physics, University of Bras\'ilia, 70910-900, Bras\'ilia, Brazil}
\email{ribeirojr@unb.br}
\author{Pulickel M. Ajayan}
\affiliation{Materials Science and Nanoengineering, Rice University, Houston, Texas 77005, USA.}
\author{Douglas S. Galvão}
\affiliation{Group of Organic Solids and New Materials (GSONM), Gleb Wataghin Institute of Physics, University of Campinas (UNICAMP), Campinas, SP, Brazil.}
\alsoaffiliation{Center for Computational Engineering \& Sciences (CCES), University of Campinas - UNICAMP, Campinas, SP, Brazil.}
\email{galvao@ifi.unicamp.br}

\title[An \textsf{achemso} demo]
  {Structural Stability of Sulfur Depleted MoS\textsubscript{2}}

\abbreviations{IR,NMR,UV}

\begin{document}

\begin{tocentry}

\includegraphics[width=0.75\linewidth]{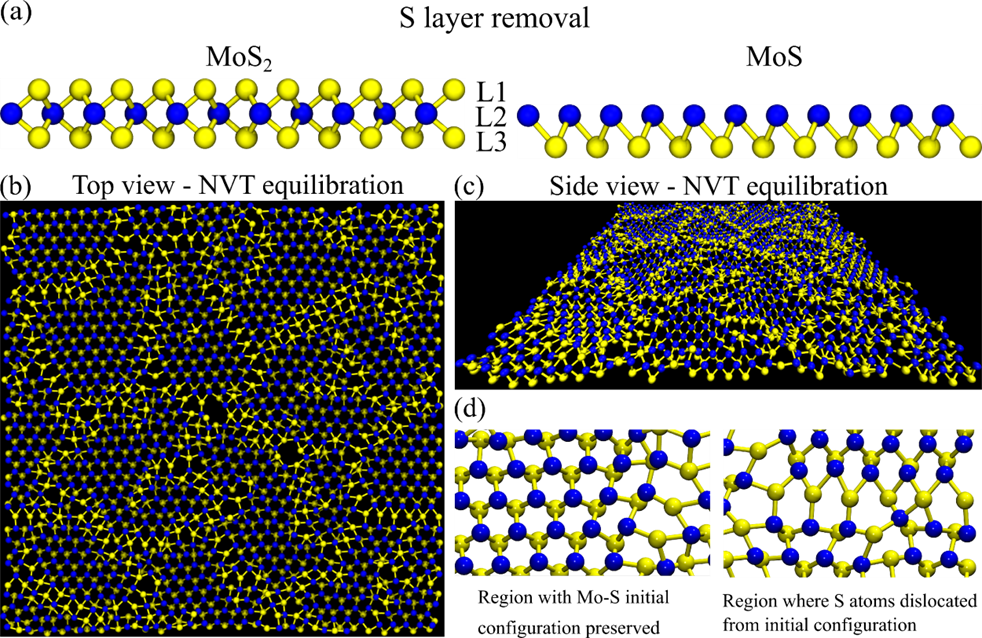}

Sulfur depleted MoS\textsubscript{2} sheet.

\end{tocentry}

\begin{abstract}
Transition metal dichalcogenides (TMDs), particularly monolayer MoS\textsubscript{2}, have received increased attention in materials science and have been exploited in diverse applications from photonics to catalysis. Defects in TMDs play a crucial role in modulating their properties, and understanding defect-induced dynamics is of great importance. This study investigates the dynamics of sulfur depletion in defective monolayer MoS\textsubscript{2}, which yields stable MoS monolayers. Various defect sizes, temperature regimes, and substrate effects were investigated. Through comprehensive classical molecular (ReaxFF) molecular dynamics (MD) and \textit{ab initio} MD (AIMD) simulations, we elucidate the dynamics of sulfur vacancy formation in MoS\textsubscript{2} lattices. After removing all sulfur atoms from the top layer, several sulfur atoms from the bottom layer spontaneously migrate to the top layer as a response to increase structural stability, thus creating a MoSx alloy. These findings deepen our understanding of defect dynamics in TMDs, offering valuable insights into the controlled engineering of their properties for nanotechnology applications.
\end{abstract}

\section{Introduction}

Transition metal dichalcogenides (TMDs) have recently received significant attention due to their attractive properties, impacting various research fields, including photonics, plasmonics, valleytronics, catalysis, optoelectronics, and flexible electronics\cite{xia2014two,mak2016photonics,agarwal2018plasmonics,schaibley2016valleytronics,deng2016catalysis,asadi2016nanostructured,zhang2017janus,fiori2014electronics,wang2012electronics,jariwala2014emerging,pospischil2014solar}. Monolayer TMDs consist of three atomic layers: a hexagonally close-packed transitional metal layer sandwiched between two chalcogen layers, forming an MX\textsubscript{2}-type compound (M = Mo, W, Sn, Nb, Ta, etc.; X = S, Se, Te)\cite{lee2012synthesis,apte2018telluride}. Among the 2D materials family, MoS\textsubscript{2} is of particular interest, having been the object of several studies due to its electrical\cite{radisavljevic2011single}, optical\cite{bernardi2013extraordinary}, and other physical properties, from bulk to monolayer forms\cite{zeng2012valley,cai2015vacancy}. 

Defects in the structure of these materials can significantly impact their properties \cite{lin2016defect}. In the case of MoS\textsubscript{2}, sulfur vacancies induce n-doping, while molybdenum ones induce p-type behavior\cite{mcdonnell2014defect}. Structural defects have also been shown to affect mobility and device performance significantly \cite{ochedowski2013radiation,wu2016defects}. Generally, changes in the electronic structure caused by defects lead to changes in excitonic processes, thereby modifying the optical properties. For instance, the photoluminescence (PL) spectrum of MoSe\textsubscript{2} exhibits shifts in its peaks with the creation of Se vacancies \cite{mahjouri2016tailoring}. On the other hand, new peaks appear in the PL spectrum of MoS\textsubscript{2} when bi-sulfur vacancies are created through $\alpha$ particle bombardment\cite{tongay2013defects}.

Proton irradiation of MoS\textsubscript{2} samples has been found to induce ferromagnetic behavior, likely due to the generation of vacancies and the contribution from defective zigzag or armchair edges that create surface states\cite{mathew2012magnetism}. Consequently, it is crucial to create defects in a controlled manner to tailor material properties. Methods for achieving this control include bombardment with argon ions\cite{ma2013controlled}, He ions\cite{stanford2016focused,fox2015nanopatterning}, $\alpha$ particles\cite{tongay2013defects}, protons\cite{mathew2012magnetism}, Mn\textsuperscript{2+} ions\cite{mignuzzi2015effect}, exposure to oxygen plasma\cite{ye2016defects}, and electron beam irradiation\cite{zan2013control}, among others.

One application where defects in TMDs play a crucial role is the hydrogen evolution reaction (HER). TMDs are well-known for their chemically inert basal planes, necessitating the introduction of defects to enhance reactivity in these materials~\cite{tsai2017electrochemical}. Theoretical calculations have revealed that sulfur vacancies in MoS\textsubscript{2} expose Mo atoms and their d-orbitals, making them more reactive\cite{le2014single}. Experimental studies have confirmed that S-vacancies in 2H-MoS\textsubscript{2} exhibit higher activity for HER compared to edge sites or the 1T-phase \cite{tsai2017electrochemical,li2016kinetic}. Furthermore, it has been demonstrated that increasing the number of vacancies enhances the reaction efficiency, and applying strain to the defective material further boosts HER activity\cite{li2016activating}.

Highly defective TMD structures can also serve as crucial post-synthesis modifications for rationalizing 2D alloys. Literature has previously demonstrated that sulfur atoms in the top layer of MoS\textsubscript{2} can be selectively sputtered by Ar$^+$ ions, followed by Se substitution through Se evaporation, leading to the synthesis of the alloy MoS$_{2(1-x)}$Se$_{2x}$ \cite{ma2014postgrowth}. Additionally, Ghorbani et al. discussed the possibility of synthesizing an alloy, MoSF, composed of molybdenum, sulfur, and fluorine, exhibiting metallic behavior through a similar process\cite{ghorbani2017two}.

In this work, we have assessed the stability of a highly defective MoS\textsubscript{2} structure through classical molecular dynamics (MD) and \textit{ab initio} molecular dynamics (AIMD) simulations. This defective lattice is obtained by systematically removing the entire top sulfur layer to examine the structural rearrangements of the resulting configuration at room temperature. Following the removal of half of the sulfur atoms, the resulting structure is afterward referred to as MoS, maintaining stoichiometric consistency. Additionally, we have investigated the interaction between the equilibrated MoS layer and a pristine MoS\textsubscript{2} one. Subsequently, the interaction was analyzed to investigate how it evolves with increasing temperatures, from 300 K to 1000 K.

\section{Results}

\begin{figure}[htbp]
    \centering
    \includegraphics[width=\linewidth]{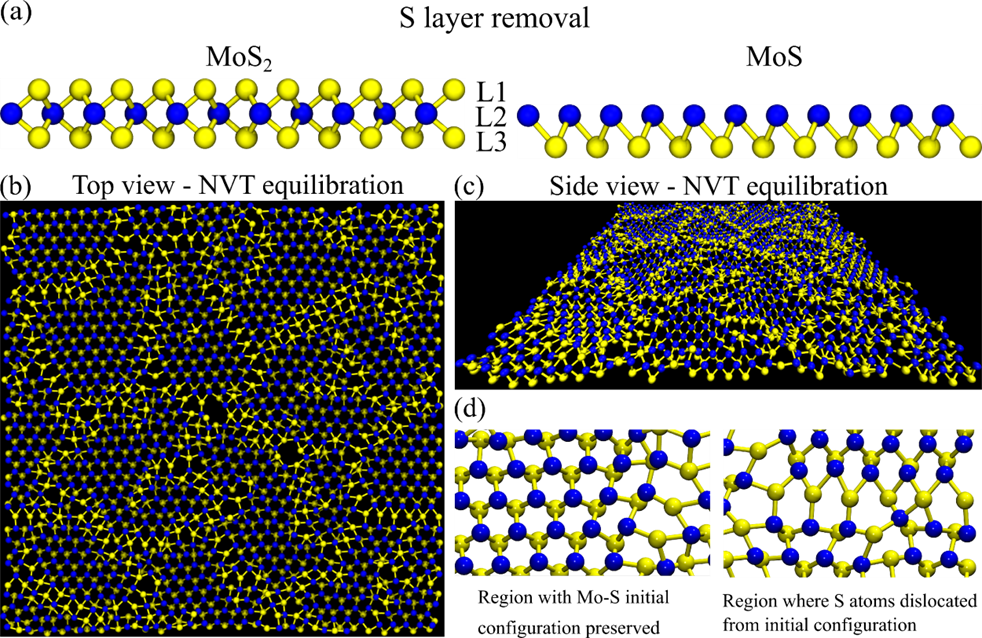}
    \caption{(a) A schematic representation of a MoS\textsubscript{2} monolayer with one of its sulfur layers (L1) removed, leading to the formation of a MOS structure containing only one Mo layer (L2) and one S layer (L3). (b) Top and (c) side views of equilibrated MoS configuration after 100 ps of classical MD in the NVT ensemble. (d) Detailed view of two regions within the MoS structure, illustrating areas where the initial MoS configuration was preserved and areas displaying significant S dislocations.}
    \label{fig1}
\end{figure}

Figure \ref{fig1}(a) illustrates the schematic representation of the MoS\textsubscript{2} monolayer configuration considered in this study. Initially, we began with a pristine MoS\textsubscript{2} structure measuring approximately 10 $\times$ 10 nm. One of its sulfur layers (L1) was entirely removed to create the MoS structure. Energy minimization calculations were carried out for both MoS\textsubscript{2} and MoS structures, resulting in relaxed Mo-S bond lengths of approximately 0.243 nm and 0.279 nm, respectively. The removal of L1 in MoS\textsubscript{2}led to an increase in the bond length values. Still, all remaining sulfur atoms remain in the L3 layer, indicating that this configuration is, at the very least, a local energy minimum.

After energy minimization, we carried out molecular dynamics simulations of MoS in an NVT ensemble at 300 K. The structure was equilibrated within 100 ps. Within approximately 10 ps, sulfur atoms began migrating to the other side of the Mo layer (L2), which, consequently,  has been depleted of sulfur. Figures \ref{fig1}(b) and \ref{fig1}(c) illustrate that this migration was incomplete but occurred as line dislocations scattered across the surface. In the regions between these lines of sulfur atoms, it was observed that large structure areas retained the initial MoS configuration, as depicted in Figure \ref{fig1}(d). Importantly, the Supplementary Material shows MD videos for the sulfur depletion process in defective monolayer MoS\textsubscript{2}. These videos capture the sulfur depletion dynamics over 20 ps in a monolayer MoS\textsubscript{2} with dimensions of 100 nm $\times$ 100 nm, containing 236800 atoms. Videos 1, 2, and 3 show different perspectives: the full view, a zoomed-in view, and a lateral zoomed-in view, respectively. In these videos, we can observe that the MoS formation still holds for large systems.  

\begin{figure}[htbp]
    \centering
    \includegraphics[width=\linewidth]{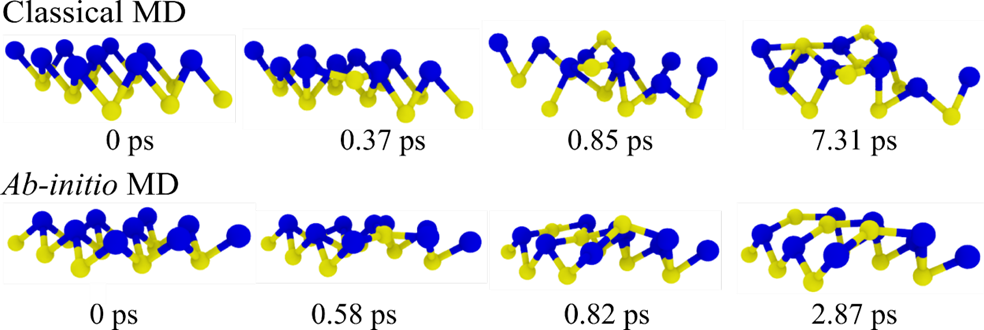}
    \caption{Dynamics of a $3\times 3$ supercell of MoS for classical and \textit{ab initio} MD simulations. Both systems reached equilibrium within a few picoseconds, during which sulfur (S) atoms transverse through the Mo layer.}
    \label{fig2}
\end{figure}

A similar process occurs in smaller systems, allowing us to contrast the system dynamics using classical and ab initio MD. Figure \ref{fig2} presents the time evolution of a MoS structure, revealing that sulfur atoms transverse the Mo layer within a few ps of dynamics, after which the system remains stable. Importantly, consistently, for both methodologies, a single S atom moves first, followed by two others. At the same time, the remaining S atoms stay at the L3 layer.

\begin{figure}[htbp]
    \centering
    \includegraphics[width=\linewidth]{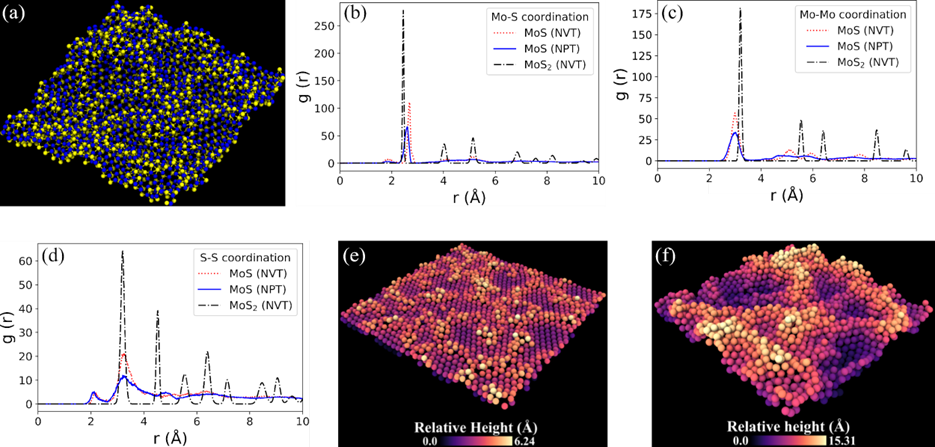}
    \caption{(a) Equilibrated MoS configuration after 100 ps of classical MD in the NPT ensemble. Coordination analysis between (b) Mo-S, (c) Mo-Mo, and (d) S-S for MoS and MoS\textsubscript{2} monolayers. Relative height mapping of MoS after equilibration in (e) NVT and (f) NPT ensembles.}
    \label{fig3}
\end{figure}

To confirm the system's stability under conditions where the simulation box dimensions were allowed to change along the $x$ and $y$ directions (i.e., along the surface), we performed additional classical molecular dynamics (MD) simulations in an NPT ensemble using a larger system. Figure \ref{fig3}(a) illustrates that the monolayer undergoes a rapid structural contraction, resulting in regions where the initial Mo-S configuration is preserved but becomes partially curved. This structural shrinking process occurs within approximately 30 picoseconds, and after 100 picoseconds, the system stabilizes.

Figure \ref{fig3}(b) presents the coordination of Mo-S atoms in both MoS and MoS\textsubscript{2} structures. Notably, the first peak in the coordination function for MoS\textsubscript{2} is well-defined at approximately 0.243 nm. In contrast, this distance exhibits a broader distribution in MoS (for both NVT and NPT ensembles). This broadening occurs in association with changes in the Mo-Mo coordination, as depicted in Figure \ref{fig3}(c). In the case of MoS, the Mo-Mo coordination pattern widens and displays nearest-neighbor peaks smaller than those in pristine MoS\textsubscript{2}. This phenomenon results from sulfur depletion on the surface, leading to a decrease in the Mo-Mo distance. This reduced distance allows sulfur atoms to migrate across the surface at locations where adjacent Mo atoms are separated. As sulfur atoms in that region move toward the depleted area (L1), the adjoining parts expand, promoting the migration of more sulfur atoms across the Mo layer.

As the system evolves, an equilibrium is achieved when a sufficient number of S atoms migrate across the surface and form Mo clusters. This process leads to a substantial change in the S-S coordination in MoS (see Figure \ref{fig3}(d)), contrasting with the behavior in MoS\textsubscript{2}. Notably, boundary lines of S atoms emerge when they transverse the Mo layer, resulting in an S-S coordination of approximately 0.2 nm. The S-S coordination widens for the S atoms that remain in the L3 layer compared to MoS\textsubscript{2}. However, the peak location remains roughly the same.

It was also observed that during the NVT equilibration, the monolayer undergoes significant changes, resulting in a corrugated surface with a thickness of approximately 0.624 nm (see Figure \ref{fig3}(e)). This corrugation becomes even more pronounced when the system is switched to the NPT ensemble, as evident in the height map shown in Figure \ref{fig3}(f). Moreover, this trend suggests that, despite achieving stability in terms of constant area (in the NVT ensemble), the extent of sulfur (S) depletion is substantial enough that the system continues to strive to lower its energy by reducing the overall distance between atoms.

Until now, our investigation has focused on removing one layer of sulfur atoms from a suspended MoS\textsubscript{2} monolayer. In experimental settings, TMDs are typically synthesized on substrates, and this interaction can significantly alter the material's dynamics. The presence of a substrate restrains movement normal to the surface. It also introduces surface interactions, both of which can be fundamental in determining the final morphology of defective TMDs.

\begin{figure}[htbp]
    \centering
    \includegraphics[width=\linewidth]{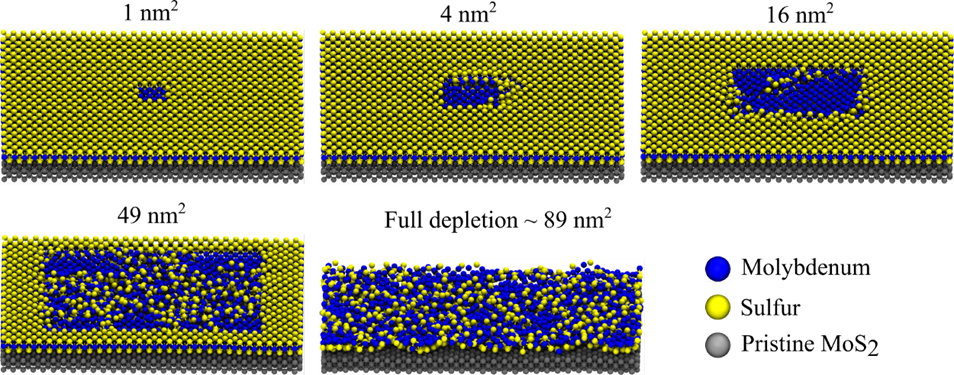}
    \caption{Equilibrated configurations after NPT runs for systems with varying defect sizes, ranging from 1 to 89 nm\textsuperscript{2} on top of a pristine MoS\textsubscript{2} substrate.}
    \label{fig4}
\end{figure}

To investigate these substrate effects, we have carried out additional simulations using systems where our MoS monolayer was placed atop pristine MoS\textsubscript{2}, effectively mimicking the role of a substrate. We systematically varied the size of the defect area in these simulations, as illustrated in Figure \ref{fig4}.

For the smallest defect area considered here (1 nm\textsuperscript{2}), we have observed an intriguing phenomenon where no sulfur atoms from L3 moved to L1. The depletion of sulfur atoms caused the Mo atoms in the defect area to approach each other, resulting in a local shrinkage of the structure. This process formed a slight indentation on the surface, preventing the sulfur atoms underneath from migrating to the L1 site. This phenomenon bears similarities to the effect of chalcogenide substitution in Janus structures, where, for instance, selenium atoms replacing sulfur on one side of MoS\textsubscript{2} can lead to the formation of scrolls and/or nanotubes, as demonstrated in reference~\cite{C8CP02011F}.

Regarding a defect area of 4 nm\textsuperscript{2}, we have observed a similar effect where the local contraction of Mo atoms caused the boundary between MoS and MoS\textsubscript{2} to move farther apart. This expansion provided sufficient space for S atoms to migrate from L3 to L1. Notably, this phenomenon was also observed as the defect area increased (49 nm\textsuperscript{2} and 89 nm\textsuperscript{2}). As the defect area grew, it increasingly resembled the total S removal shown previously in Figure \ref{fig1}. The appearance of complete S depletion on top of the MoS\textsubscript{2} substrate is like that of suspended MoS. However, the relative height of MoS in this substrate-supported scenario, measured at 5.77 \r{A} after NPT equilibration, was even smaller than in the suspended case during NVT equilibration. This trend indicates that a substrate has a flattening effect on the MoS surface due to interlayer interactions. Nevertheless, the dynamics of S atoms moving from L3 to L1 remained consistent if the defect area was sufficiently large (approximately 4 nm\textsuperscript{2}).

In our simulations carried out at 300 K, we have observed single Mo-S bonds and no bonding between the Mo atoms from MoS and the S atoms of the pristine MoS\textsubscript{2} substrate. However, to explore how this system behaves at elevated temperatures, we carried out further simulations of the MoS/MoS\textsubscript{2} system at temperatures of 600, 800, and 1000 K. Upon increasing the temperature to 600 K, we observed an increase in bondings between the Mo atoms from MoS and the S atoms of MoS\textsubscript{2}. This phenomenon arises because the Mo atoms in the MoS layer, where S has migrated from L3 to L1, become more mobile. Higher temperatures provide them the energy required to overcome the thermal barriers and to interact with the S atoms from MoS\textsubscript{2}.

\begin{figure}[htbp]
    \centering
    \includegraphics[width=\linewidth]{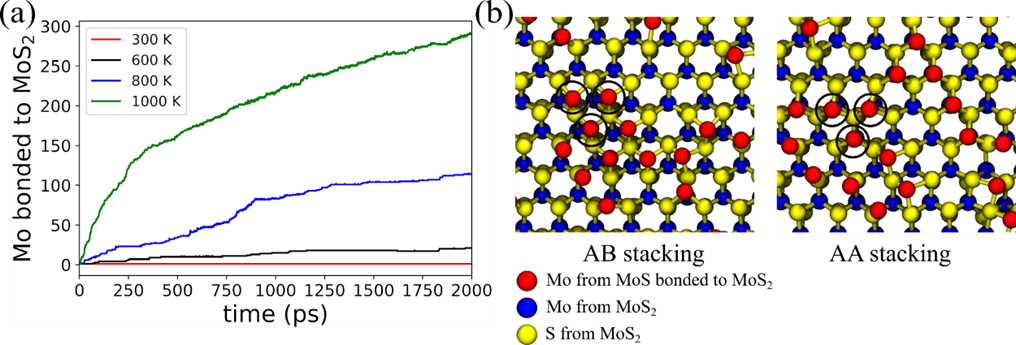}
    \caption{(a) The number of Mo atoms from MoS added to the MoS\textsubscript{2} layer. (b) MD snapshots showing AB and AA stacking configurations formed between pristine MoS\textsubscript{2} and Mo from MoS after 2 ns of equilibration at 1000 K.}
    \label{fig5}
\end{figure}

As the temperature rises, the number of Mo atoms from MoS that are adsorbed onto MoS\textsubscript{2} also increases, as illustrated in Figure \ref{fig5}(a). Higher temperatures accelerate the process of Mo adsorption, resulting in the faster degradation of the MoS layer. However, up to the 2 ns timescale, the system had not yet reached complete adsorption. Another interesting fact is that the coordination of Mo atoms added to the pristine MoS\textsubscript{2} structure follows AB and AA stacking formations along the surface. The regions are clustered in either of the stacked configurations, as some Mo atoms are positioned on top of S atoms.

\section{Computational Details}

In this study, we have carried out comprehensive fully atomistic molecular dynamics (MD) simulations, employing both classical and ab initio approaches to investigate the behavior of defective MoS2 monolayers and bilayers. For the classical MD simulations, we used the LAMMPS code \cite{plimpton1995fast} within the framework of the reactive ReaxFF potential \cite{chenoweth2008reaxff}. Our systems encompassed a 3 $\times$ 3 $\times$ 1 hexagonal supercell for the MoS2 monolayer, characterized by dimensions of 9.57 $\times$ 9.57 $\times$ 43.13 Å. For the larger monolayer and bilayer MoS sheets, we have employed an orthorhombic simulation box measuring approximately 95.69 $\times$ 93.92 $\times$ 50 Å.

We started our simulations using a conjugate gradient algorithm to minimize the energy of pristine MoS2 systems. Subsequently, we selectively removed S atoms, including the top sulfur layer, in both monolayer and bilayer configurations. Afterward, the systems underwent energy minimization, followed by equilibration in the canonical ensemble at 300 K for 100 ps and the isobaric ensemble at 300 K and 0 bar.

In the bilayer setup, we restrained S atoms at the bottom of the pristine layer with a force of 10 kcal/mol along the z-direction to mimic substrate effects. Temperature and pressure were controlled using a Nosé-Hoover thermostat \cite{nose1984unified,hoover1985canonical}, with time constants of 10 fs for temperature and 100 fs for pressure regulation.

For the \textit{ab initio} MD simulations, we have used the Quantum Espresso code \cite{giannozzi2017advanced,giannozzi2009quantum}. Our study focused on hexagonal supercells composed of 3 $\times$ 3 $\times$ 1 unit cells of monolayer MoS2, consistent with our previous classical MD simulations. Before considering the cases involving sulfur-depleted structures, we first relaxed the pristine configurations. Electron exchange-correlation effects were treated using the Perdew–Burke–Ernzerhof (PBE) functional within the generalized gradient approximation (GGA). The Kohn-Sham equations were solved using the projector augmented wave (PAW) method with a plane wave basis set and a cutoff energy of 500 Ry.

In our structural relaxation simulations, we have used a Monkhorst–Pack scheme for k-point sampling, utilizing a 14$\times$14$\times$1 k-point mesh. However, for the ab initio MD calculations, we opted for a more efficient approach by using only the gamma point to save substantial computational time. We have used the Verlet integrator, allowing the simulation box to move/relax along the surface in the x and y dimensions while keeping the z dimension fixed. Structures were relaxed with convergence thresholds of $10^{-4}$ Ry/Bohr for atomic forces on each ion and $10^{-6}$ Ry for electronic relaxation. We introduced a 20 Å vacuum slab along the z-direction to prevent spurious interactions between periodic (mirror) images.

\section{Conclusions}

In summary, our study has revealed the remarkable stability of monolayer MoS2 structures with one complete sulfur layer removed, aptly referred to as MoS. After removing all sulfur atoms from the top layer, several sulfur atoms from the bottom layer spontaneously migrate to the top layer as a response to increase structural stability, thus creating a MoSx alloy. Our findings are based on consistent results obtained through two distinct methodologies: ab initio and classical reactive molecular dynamics. Suspended and supported MoS monolayers exhibit stability over large areas, forming extensive regions where the initial Mo-S structure remains intact. This stability, however, comes at the cost of some sulfur atom dislocation to the depleted region of the monolayer.

Our simulations unveil an interesting temperature-dependent phenomenon in the context of supported MoS monolayers. At elevated temperatures, Mo atoms from MoS bond with the underlying pristine MoS2 layer, forming islands with AB or AA stacking configurations. These insights into atomic manipulation within TMDs offer a promising avenue for creating novel morphologies and synthesizing Janus structures, paving the way for innovative applications of these compounds in nanoscience and technology. This investigation enhances the understanding of defect dynamics in TMDs and provides practical approaches to engineering their properties for diverse nanotechnology applications.

\begin{acknowledgement}

This work was financed in part by the Coordenação de Aperfeiçoamento de Pessoal de Nível Superior (CAPES) - Finance Code 001 and grant 88887.691997/2022-00, Conselho Nacional de Desenvolvimento Científico e Tecnológico (CNPq), FAP-DF, and FAPESP. M.L.P.J. acknowledges financial support from FAPDF (grant 00193-00001807/2023-16), CNPq (grant 444921/2024-9), and CAPES (grant 88887.005164/2024-00). M.L.P.J and L.A.R.J thank also to CENAPAD-SP (National High-Performance Center in São Paulo, State University of Campinas -- UNICAMP, projects: proj960 and proj634) and NACAD (High-Performance Computing Center, Lobo Carneiro Supercomputer, Federal University of Rio de Janeiro -- UFRJ, projects: a22002 and a22003) for the computational support provided. We thank the Center for Computing in Engineering and Sciences at Unicamp for financial support through the FAPESP/CEPID Grants \#2013/08293-7 and \#2018/11352-7. L.A.R.J acknowledges the financial support from FAP-DF grant 00193-00001857/2023-95, FAPDF-PRONEM grant 00193.00001247/2021-20, and CNPq grants 302922/2021-0 and 167745/2023-9. 
\end{acknowledgement}

\begin{suppinfo}
Molecular dynamics videos illustrate the sulfur depletion process in defective monolayer MoS\textsubscript{2}. Videos 1, 2, and 3 show different perspectives: the full view, a zoomed-in view, and a lateral zoomed-in view, respectively. These videos capture the sulfur depletion dynamics over 20 ps in a monolayer MoS\textsubscript{2} with dimensions of 100 nm $\times$ 100 nm, approximately, containing 236800 atoms.
\end{suppinfo}
\bibliography{bibliography}

\end{document}